\documentclass[12pt,preprint]{aastex}

\slugcomment{}

\shorttitle{High Latitude Molecular Clouds in Pegasus}
\shortauthors{Yamamoto et al.}

\begin{document}

\title{Large Scale CO Observations of a Far-Infrared Loop in Pegasus; Detection of a Large Number of Very Small Molecular Clouds Possibly Formed via Shocks}

\author{H.\ Yamamoto\altaffilmark{1}, A.\ Kawamura\altaffilmark{1}, K.\ Tachihara\altaffilmark{2}, N.\ Mizuno\altaffilmark{1}, T.\ Onishi\altaffilmark{1} and Y.\ Fukui\altaffilmark{1}}

\altaffiltext{1}{Department of Astrophysics, Nagoya University, Chikusa-ku, Nagoya, Japan 464-8602; hiro@a.phys.nagoya-u.ac.jp}
\altaffiltext{2}{Graduate School of Science and Technology, Kobe University, 1-1 Rokko-dai, Nada-ku, Kobe, Japan 657-8501}

\begin{abstract}

We have carried out large scale CO observations with a mm/sub-mm telescope NANTEN toward a far infrared loop-like structure whose angular extent is about 20$\times$20 degrees around ($l$, $b$) $\sim$ (109$\degr$, $-$45$\degr$) in Pegasus. Its diameter corresponds to $\sim$ 25 pc at a distance of 100 pc, adopted from that of a star HD886 (B2IV) near the center of the loop.  We covered the loop-like structure in the $^{12}$CO ($J$ = 1--0) emission at 4$\arcmin$--8$\arcmin$ grid spacing and in the $^{13}$CO ($J$ = 1--0) emission at 2$\arcmin$ grid spacing for the $^{12}$CO emitting regions.  The $^{12}$CO distribution is found to consist of 78 small clumpy clouds whose masses range from 0.04 $M_{\sun}$ to 11 $M_{\sun}$, and $\sim$ 83\% of the $^{12}$CO clouds have very small masses less than 1.0 $M_{\sun}$.   $^{13}$CO observations revealed that 18 of the 78 $^{12}$CO clouds show significant $^{13}$CO emission. 
$^{13}$CO emission was detected in the region where the column density of H$_{2}$ derived from $^{12}$CO is greater than 5$\times$10$^{20}$ cm$^{-2}$, corresponding to $A$v of $\sim$ 1 mag, which takes into account that of H{\small \,I}.  We find no indication of star formation in these clouds in IRAS Point Source Catalog and 2MASS Point Source Catalog. The very low mass clouds, $M$ $\leq$ 1$M_{\sun}$, identified are unusual in the sense that they have very weak $^{12}$CO peak temperature of 0.5 K--2.7 K and that they aggregate in a region of a few pc with no main massive clouds; contrarily to this, similar low mass clouds less than 1 $M_{\sun}$ in other regions previously observed including those at high Galactic latitude are all associated with more massive main clouds of $\sim$ 100 $M_{\sun}$. A comparison with a theoretical work on molecular cloud formation (Koyama \& Inutsuka 2002) suggests that the very low-mass clouds may have been formed in the shocked layer through the thermal instability. The star HD886 (B2IV) may be the source of the mechanical luminosity via stellar winds to create shocks, forming the loop-like structure where the very low-mass clouds are embedded.
\end{abstract}


\section{INTRODUCTION}

 High Galactic latitude molecular clouds (hereafter HLCs) are typically located at $\mid$b$\mid$ $\gtrsim$ 20$\degr$--30$\degr$.  Since the Gaussiun scale height of CO is estimated to be $\sim$ 100 pc in the inner Galactic disk  (e.g., Magnani et al. 2000), HLCs are likely located very close to the Sun, within a few hundred pc or less.  Their proximity to the Sun and the low possibility of overlapping with other objects along the line of sight enable us to study them with a high spatial resolution and to compare CO data unambiguously with the data at other wavelengths.  HLCs have lower molecular densities compared with dark clouds where the optical obscuration is significant.  Therefore, HLCs are often called as translucent clouds (e.g., van Dishoeck \& Black 1988) and most of the known HLCs are not the sites of active star formation, although a few of them are known to be associated with T Tauri stars (e.g., Magnani et al. 1995; Pound 1996; Hearty et al. 1999).

Given the very small distances of HLCs, it is a challenging task for observers to make a complete survey for HLCs over a significant portion of the whole sky. $^{12}$CO ($J$ = 1--0) emission has been used to search for HLCs because the line emission in the mm band is strongest among the thermally or sub-thermally excited spectral lines of interstellar molecular species.  It is however difficult to cover an area as large as tens of square degrees subtended by some of the HLCs because of the general weakness of the $^{12}$CO emission, typically $\sim$ a few K (e.g., Magnani et al. 1996), with existing mm-wave telescopes in a reasonable time scale.  HLCs have been therefore searched for by employing various large-scale datasets at other wavelengths including the optical obscuration (Magnani et al. 1985; Keto \& Myers 1986), the infrared radiation (Reach et al. 1994), and the far-infrared excess over H{\small \,I} (=FIR excess)(Blitz et al. 1990; Onishi et al. 2001).  On the other hand, unbiased surveys in CO at high Galactic latitudes have been performed at very coarse grid separations of 1$\degr$ resulting in a small sampling factor of a few \% (Hartmann et al. 1998; Magnani et al. 2000).  Most recently, Onishi et al. (2001) discovered 32 HLCs or HLC complexes. This search was made based on the FIR excess, demonstrating the correlation among FIR excess clouds with CO clouds is a useful indicator of CO HLCs.

Previous CO observations of individual HLCs at higher angular resolutions show that HLCs exhibit often loop-like or shell-like distributions having filamentary features with widths of several arc min or less (Hartmann et al. 1998; Magnani et al. 2000; Bhatt 2000), and in addition that HLCs often compose a group, whose angular extent is $\sim$ 10 degrees or larger.  In order to better understand the structure of HLCs and to pursue the evolution of HLC complexes, CO observations covering tens of square degrees at a high angular resolution are therefore crucial.  The past observations of such complexes of HLCs are limited to a few regions including Polaris flare (Heithausen \& Thaddeus 1990), Ursa Major (Pound \& Goodman 1997) and the HLC complex toward MBM 53, 54, and 55 (Yamamoto et al. 2003).  Pound \& Goodman (1997) showed an arc-like structure of the molecular cloud system and suggested that the origin of such structures could be some explosive events.  Most recently, Yamamoto et al. (2003) carried out extensive observations of the molecular cloud complex including MBM 53, 54, and 55 and suggest that the HLCs may be significantly affected by past explosive events based on the arc-like morphologies of molecular hydrogen (see also Gir et al. 1994).

The region of MBM 53, 54, and 55 is of particular interest among the three, because it is associated with a large H{\small \,I} cloud of $\sim$ 590 $M_{\sun}$ at a latitude of $-$35 degrees and because there is a newly discovered HLC of 330$M_{\sun}$, HLCG92$-$35, which is significantly H{\small \,I} rich with a mass ratio $M$(H$_{2}$)/$M$(H{\small \,I}) of $\sim$ 1,  among the known HLCs (Yamamoto et al. 2003).  This cloud was in fact missed in the previous surveys based on optical extinction (Magnani et al. 1985). Subsequent to these observations we became aware of that the region is also very rich in interstellar matter as shown by the 100$\mu$m dust features (Kiss et al. 2004). There is a loop-like structure shown at 100 $\mu$m around ($l$, $b$) $\sim$ (109$\degr$, $-$45$\degr$).  Toward the center of the loop, an early type star HD886(B2IV) is located and may play a role in creating the loop. Its proper motion is large at a velocity of a few km s$^{-1}$, suggesting that the stellar winds of the star might have continued to interact with the surrounding neutral matter over a few tens of pc in $\sim$ a few Myr. Magnani et al. (1985) and Onishi et al. (2001) yet observed only a small part of this region. In order to reveal the large scale CO distribution of the region, we have carried out observations toward ($l$, $b$) $\sim$ (109$\degr$, $-$45$\degr$) by $^{12}$CO ($J$ = 1--0) and $^{13}$CO ($J$ = 1--0) with NANTEN 4-meter millimeter/sub-mm telescope of Nagoya University at Las Campanas, Chile. We shall adopt the distance of 100 pc from the sun to the loop-like structure which is equal to the distance of the B2 star in the center of the loop, and is also a typical value for the HLCs.

\section{OBSERVATIONS}

  $^{12}$CO ($J$ = 1--0) and $^{13}$CO ($J$ = 1--0) observations were made with the 4-meter telescope, NANTEN, of Nagoya University at Las Campanas Observatory of Carnegie Institutions of Washington, Chile.  The front-end was an SIS receiver cooled down to 4 K with a closed-cycle helium gas refrigerator (Ogawa et al. 1990).  The backend was an acousto-optical spectrometer with 2048 channels, and the total bandwidth was 40 MHz.  The frequency resolution was 35 kHz, corresponding to a velocity resolution of $\sim$ 0.1 km s$^{-1}$.  A typical system noise temperature was $\sim$ 200 K (SSB) at 115.271 GHz and $\sim$ 150 K (SSB) at 110.201 GHz.  The half-power beam width was about 2$\farcm$6, corresponding to 0.076 pc at a distance of 100 pc.  The pointing accuracy was better than 20$\arcsec$, as established by radio observations of Jupiter, Venus, and the edge of the Sun in addition to optical observations of stars with a CCD camera attached to the telescope.

 The observed region in $^{12}$CO was $\sim$ 240 square degrees toward the whole area of the loop-like structure centered at around ($l$, $b$) $\sim$ (109$\degr$, $-$45$\degr$) shown in a 100 $\mu$m map by Schlegel et al. (1998). First, the $^{12}$CO observations were made at a grid spacing of 8$\arcmin$$\times$cos($b$) and 8$\arcmin$ in Galactic longitude and latitude, respectively.  Then, the regions where the $^{12}$CO emission is significantly detected were observed at a grid spacing of 4$\arcmin$$\times$cos($b$) and 4$\arcmin$ in Galactic longitude and latitude, respectively.  The $^{13}$CO observations were made in and around the whole area where the peak temperature of $^{12}$CO emission is higher than 2.0 K at a grid spacing of 2$\arcmin$$\times$cos($b$) and 2$\arcmin$ in Galactic longitude and latitude, respectively.  The periods of $^{12}$CO observations were several sessions between 2002 May and November and those of $^{13}$CO were those between 2003 April and August.  All the observations were made by frequency switching whose interval is 20 MHz, corresponding to $\sim$ 50 km s$^{-1}$.  The integration times per point of $^{12}$CO and $^{13}$CO observations were typically $\sim$ 30 s and $\sim$ 75 s, respectively, resulting in typical rms noise temperatures per channel of $\sim$ 0.35 K and $\sim$ 0.15 K in the radiation temperature, $T_{\rm R}^*$, respectively.  In reducing the spectral data, we subtracted forth-order polynomials for the emission-free parts in order to ensure a flat spectral baseline.  Total numbers of observed points of $^{12}$CO and $^{13}$CO are 16890 and 3100, respectively.

  We employed a room-temperature blackbody radiator and the sky emission for the intensity calibration. An absolute intensity calibration and the overall check of the whole system were made by observing Orion KL [$\alpha$(1950) = 5$^{\rm h}$32$^{\rm m}$47.$^{\rm s}$0, $\delta$(1950) = $-$5$\degr$24$\arcmin$21$\arcsec$] every
2 hours.  We assumed the $T_{\rm R}^*$ of Orion KL to be 65 K for $^{12}$CO and 10 K for $^{13}$CO.

\section{RESULTS}

\subsection{$^{12}$CO Observation}

\subsubsection{Distribution and Past Detection of $^{12}$CO Clouds}

  Figure 1 shows the distribution of the velocity-integrated intensity map of $^{12}$CO emission.  We defined a $^{12}$CO cloud as a collection of more than two contiguous observed positions whose integrated intensity exceeds 0.77 K km s$^{-1}$ (5$\sigma$). Based on the definition, we identified 78 molecular clouds in this region. Molecular clouds are concentrated from ($l$, $b$) $\sim$ (107$\degr$, $-$37$\degr$) to (116$\degr$, $-$45$\degr$) and around ($l$, $b$) $\sim$ (114$\degr$, $-$52$\degr$). Most of the molecular clouds are very small, having size of $\lesssim$ 1$\degr$. Figure 2 shows the distribution of the CO superposed on the SFD 100 $\mu$m (Schlegel et al. 1998), which was derived from a composite of the COBE/DIRBE and IRAS/ISSA maps, with the foreground zodiacal light and confirmed point sources removed. CO clouds are distributed along the infrared loop whose diameter is $\sim$ 25 pc. We detected little CO emission within the loop-like structure, while toward some of the local peaks of SFD 100 $\mu$m there is no CO emission.

Figure 3 shows the peak radial velocity distribution derived from the present $^{12}$CO data set. The velocity in Figure 3 is derived by a single gaussiun fitting from all CO spectra. 
The velocity range of the molecular clouds is from $-$18.3 km s$^{-1}$ to 0.3 km s$^{-1}$ and there is no systematic large scale velocity gradients.

  Some of the molecular clouds have already been known by previous observations. Molecular clouds toward ($l$, $b$) $\sim$ (110$\fdg$18, $-$41$\fdg$23) and (117$\fdg$36, $-$52$\fdg$28) are identified by Magnani et al. (1985) and named as MBM 1 and MBM 2, respectively. DIR117$-$44 and DIR105$-$38 identified by Reach et al. (1998) are also identified in CO toward ($l$, $b$) $\sim$ (116$\fdg$5, $-$44$\fdg$0) and (105$\fdg$0, $-$38$\fdg$0) by Onishi et al. (2001). Magnani et al. (1986) detected CO emission at ($l$, $b$) $\sim$ (112$\degr$, $-$40$\degr$). Magnani et al. (2000) also covered this region even though they made observations on a locally Cartesian grid with 1$\degr$(true angle) spacing in longitude and latitude for a beam size of 8$\farcm$8, they detected CO emission at eight positions of ($l$, $b$) $\sim$ (103$\fdg$2, $-$38$\fdg$0), (103$\fdg$2, $-$39$\fdg$0), (104$\fdg$4, $-$39$\fdg$0), (106$\fdg$8, $-$37$\fdg$0), (108$\fdg$0, $-$52$\fdg$0), (109$\fdg$5, $-$51$\fdg$0), (110$\fdg$4, $-$41$\fdg$0), and (111$\fdg$0, $-$50$\fdg$0) in the present region while they missed the present small molecular clouds whose sizes are less than several arc min in Figure 1 due to the coarse grid spacing.

\subsubsection{Physical Properties of $^{12}$CO Molecular Clouds}

  Seventy-eight $^{12}$CO molecular clouds are identified in the present region. For each molecular cloud, $\Delta V$ derived from single Gaussian fitting was from 0.5 to 3.7 km s$^{-1}$, and the radial velocity, $V_{\rm LSR}$, ranges from $-$15.7 to $-$0.1 km s$^{-1}$.  The maximum brightness temperature, $T_{\rm R}^*$($^{12}$CO) ranges from 0.5 to 5.7 K.  The radius of a cloud, $R$, which is defined as the radius of an equivalent circle having the same area, i.e., and $R$(pc)$=\sqrt{A/\pi}$ where A is the total cloud surface area within the 5$\sigma$--contour level, ranges from 0.07 to 0.79 pc.  The peak column density of molecular hydrogen, $N$(H$_{2}$), in each cloud derived by assuming a conversion factor of 1.0$\times$10$^{20}$ cm$^{-2}$/(K km s$^{-1}$) (Magnani et al. 2000) ranges from 8.0$\times$10$^{19}$ to 1.7$\times$10$^{21}$ cm$^{-2}$ with the present detection limit, 7.7$\times$10$^{19}$ cm$^{-2}$, corresponding to mass detection limit of 0.014 $M_{\sun}$. We estimate the molecular mass, $M$($^{12}$CO), by using the following formula
\begin{equation}
M(^{12}{\rm CO}) = \mu m_{{\rm H}} \Sigma[D^2 \Omega N({\rm H}_{2})],
\end{equation}
where $\mu$ is the mean molecular weight, assumed to be 2.8 by taking into account a relative helium abundance of 25\% in mass, $m_{{\rm H}}$ is the mass of the atomic hydrogen, $D$ is the distance from the Sun to the molecular clouds, and $\Omega$ is the solid angle subtended by a unit grid spacing of (4$\arcmin$)$\times$(4$\arcmin$$\times$cos($b$)).  $M$($^{12}$CO) ranges from $\sim$ 0.04 to $\sim$ 11 $M_{\sun}$ and the total mass of molecular clouds is $\sim$ 64 $M_{\sun}$.  These physical properties are listed in Table 1 and the histograms of $T_{\rm R}^*$($^{12}$CO), $\Delta V$, log($R$), and log($N$(H$_{2}$)) of these clouds are shown in Figure 4.  Histograms in Figure 4 are divided into three different categories, \textit{Usual Cloud} (hereafter UC) whose mass is greater than 1 $M_{\sun}$, \textit{Small Cloud} (hereafter SC) whose mass is  between 0.1 and 1 $M_{\sun}$, and \textit{Very Small Cloud} (hereafter VSC) whose mass is less than 0.1 $M_{\sun}$.  It is remarkable that there are a number of molecular clouds having mass less than 1 $M_{\sun}$ and that the fractions of SC and VSC are 43/78 $\sim$ 55\% and 22/78 $\sim$ 28\% in the present region, respectively. In addition, the sizes of SC and VSC are equal to or less than 0.1 pc.  We also note that the peak temperatures of SC and VSC are typically in a range from 0.5 K to 2.7 K, well below that of UC in the same region.

\subsection{The Detection and Physical Properties of the $^{13}$CO Molecular Clouds}

 Figure 5 shows the distribution of the velocity-integrated intensity map of the $^{13}$CO emission superposed on the $^{12}$CO distribution. The total area of the $^{13}$CO observations is $\sim$ 29 square degrees toward 38 of the 78 $^{12}$CO clouds. We observed all of 13 UCs, 24 of 43 SCs, and 3 of 22 VSCs.  We detected $^{13}$CO emission at 11 of the 13 UCs, 8 of the 24 SCs, and none of the 3 VSCs, indicating a trend that the $^{13}$CO intensity increases with $^{12}$CO cloud mass.

A $^{13}$CO cloud is defined in the same way as for a $^{12}$CO cloud except for the lowest integrated intensity level, 0.3 K km s$^{-1}$ (3$\sigma$).  Based on the definition, we identified 33 $^{13}$CO clouds.  For the 33 $^{13}$CO molecular clouds, $\Delta$$V$ derived from single Gaussian fitting is $\sim$ 1.5 km s$^{-1}$ and $V_{\rm LSR}$ of them ranges from $-$13.1 to $-$1.9 km s$^{-1}$.  Other physical properties, the maximum brightness temperature, $T_{\rm R}^*$($^{13}$CO), and $R$ range from 0.3 to 2.3 K and from 0.04 to 0.21 pc, respectively.  The physical parameters including  the molecular column density and mass (hereafter $M_{\rm LTE}$) are derived on the assumption of local thermodynamic equilibrium (LTE).  To derive the column density of molecular hydrogen, the optical depth of $^{13}$CO is estimated by using the following equations,
\begin{equation}
\tau(^{13}{\rm CO})={\rm ln}\left[1-\frac{T_{\rm R}^{*}(^{13}{\rm CO})}{5.29}\left\{\frac{1}{{\rm exp}(5.29/T_{\rm ex})-1}-0.164\right\}^{-1}\right],
\end{equation}
where $T_{\rm ex}$ is the excitation temperature of the $J$ = 1--0 transition of CO in K and was derived from
\begin{equation}
T_{\rm ex}=\frac{5.53}{{\rm ln}\left\{1+5.53/\left[T_{\rm R}^{*}(^{12}{\rm CO})+0.819\right]\right\}}.
\end{equation}
 $T_{\rm ex}$ was estimated to be 9.4 K from our $^{12}$CO data. The $^{13}$CO column density, $N$($^{13}$CO), is estimated by
\begin{equation}
N(^{13}{\rm CO})=2.42\times10^{14}\nonumber\times\frac{\tau(^{13}{\rm CO})T_{\rm ex}({\rm K})\Delta V({\rm km \hspace{0.1cm} s^{-1}})}{1-{\rm exp}[-5.29/T_{\rm ex}({\rm K})]}  ({\rm cm^{-2}}).
\end{equation}
The ratio of $N$(H$_{2}$)/$N$($^{13}$CO) was assumed to be 7$\times$10$^{5}$ (Dickman 1978).  The $M_{\rm LTE}$ of a cloud from $N$(H$_{2}$) is derived by the same way as $^{12}$CO (see equation (1)).  The column density and $M_{\rm LTE}$ range from 2.3$\times$10$^{20}$ to 1.7$\times$10$^{21}$ cm$^{-2}$ and 0.03 to 1.41 $M_{\sun}$, respectively, where the detection limit in the column density is 2.0$\times$10$^{20}$ cm$^{-2}$, coressponding to mass limit of 0.009 $M_{\sun}$, smaller than that of $^{12}$CO because the observations of $^{13}$CO were made by higher grid sampling and lower rms noise fluctuations than those of $^{12}$CO, respectively.  Figure 6 shows the histograms of each physical property.  The virial mass, $M_{\rm vir}$, of a cloud was derived by using the following equation, assuming isothermal, spherical, and uniform density distribution with no external magnetic pressure:
\begin{equation}
M_{\rm vir} = 209 \times R \times \Delta V_{\rm comp}^{2},
\end{equation}
where $R$ and $\Delta V_{\rm comp}$ are the radius (pc) and line width (km s$^{-1}$) of the composite profile obtained by averaging all the spectra within a cloud, respectively (for details of the line width of composite profiles, see Yonekura et al. 1997; Kawamura et al. 1998).  From this equation, $M_{\rm vir}$ is estimated to be in a range from 4.7 to 197 $M_{\sun}$.  These physical properties are also listed in Table 2.

\section{CORRELATIONS AMONG THE CLOUD PHYSICAL PARAMETERS}

\subsection{Mass Spectrum and Size Linewdith Relation}

 Figure 7a and 7b show the mass spectrum of the present  $^{12}$CO and $^{13}$CO clouds.  The spectra have been fitted by the maximum-likelihood method (Crawford et al. 1970), and it is found that they are well fitted by a single power law as follows; $dN/dM$ $\propto$ $M^{-1.53\pm0.13}$ for the $^{12}$CO clouds and $dN/dM$ $\propto$ $M^{-1.36\pm0.10}$ for the $^{13}$CO clouds.  These values of the spectral indices seem to be similar to those for the higher mass range (e.g., Yonekura et al. 1997).

 Figure 8 shows a plot of size, $R$, versus line width, $\Delta V$, of the $^{13}$CO clouds in this region and for a comparison with other HLCs, MBM 53, 54, and 55 complex (Yamamoto et al. 2003).  We can make fitting as follows by using a least-squares fitting, log($\Delta V$) = (0.22$\pm$0.43) $\times$ log($R$) + (0.37$\pm$0.52) (c.c.=0.23) for the present region and log($\Delta V$) = (0.43$\pm$0.32) $\times$ log($R$) + (0.53$\pm$0.28) (c.c.=0.37) for MBM 53, 54, and 55 complex. The low correlation coefficient (c.c.) indicates that there is no significant correlation between $R$ and $\Delta V$ because of a small range of $R$. Here we do not show the same relationship for $^{12}$CO, because the non-circular shape of the $^{12}$CO clouds may not be appropriate to derive reliable $R$.

\subsection{$M_{\rm LTE}$ vs. $M_{\rm vir}$}

 Figure 9 shows a plot of $M_{\rm LTE}$ versus $M_{\rm vir}$. The present $^{13}$CO clouds are located far above the equilibrium line where $M_{\rm LTE}$ is equal to $M_{\rm vir}$, indicating that the $^{13}$CO clouds are not in the virial equilibrium.  This indicates that none of the molecular clouds are gravitationally bound.  These parameters can be fitted by using a least-squares fitting as follows, log($M_{\rm vir}$) = (0.91$\pm$0.30) $\times$ log($M_{\rm LTE}$) + (2.23$\pm$0.29) (c.c.=0.66) for present molecular clouds and log($M_{\rm vir}$) = (0.77$\pm$0.13) $\times$ log($M_{\rm LTE}$) + (0.16$\pm$0.08) (c.c.=0.74) for MBM53, 54, and 55 complex. As mentioned in Yamamoto et al. (2003), the present molecular clouds also tend to be more virialized as the mass increases.  
For the Gemini and Auriga, and Cepheus-Cassiopeia regions, the indices of $M_{\rm LTE}$ for $M_{\rm vir}$ of $^{13}$CO clouds are estimated to be 0.72$\pm$0.03 and 0.62$\pm$0.03 for the cloud mass range of $M_{\rm LTE}$ $<$ 10$^{4}$ $M_{\sun}$ and 10$^{2}$ $M_{\sun}$ $<$ $M_{\rm LTE}$ $<$ 10$^{5}$ $M_{\sun}$, respectively (Kawamura et al. 1998; Yonekura et al. 1997).  Although the mass ranges of MBM 53, 54, and 55 complex and of this region are 10$^{-1}$ $M_{\sun}$ $<$ $M_{\rm LTE}$ $<$ 10$^{2}$ $M_{\sun}$ and 10$^{-2}$ $M_{\sun}$ $<$ $M_{\rm LTE}$ $<$ 1 $M_{\sun}$, respectively, difference in the power-law indices among these regions is small and a tendency that the SCs have large ratios of $M_{\rm vir}$/$M_{\rm LTE}$ is commonly seen.

\section{COMPARISON WITH OTHER WAVELENGTH DATA}

\subsection{No Sign of Star Formation}

In order to look for sings of star formation associated with the present molecular clouds, we searched the IRAS point source catalog for candidates of protostellar objects satisfying the following criteria:  (1) point sources having a data quality flag better than 2 in 4 bands, (2) flux ratios at 12, 25, 60 $\mu$m satisfying both log($F_{12}$/$F_{25}$) $<$ $-$0.3 and log($F_{25}$/$F_{60}$) $<$ 0, and not identified as galaxies or planetary nebulae and stars. We find that there are no cold IRAS point sources satisfying these criteria in the present molecular clouds. We also find that there are no IRAS point sources having a spectrum like a T-Tauri type star or no YSOs identified from Point Source Catalog of Two-Micron All-Sky Survey in this region. Here we select the 2MASS sources whose signal to noise ratio of valid measurements in all bands are greater than 10 and extract the sources which have the spectra like T Tauri stars in ($J$$-$$H$)--($H$$-$$K$) color--color diagram (e.g., Meyer et al. 1997) . These results suggest that the present molecular clouds are not the site of recent star-formation, or that the region is not remnants of past star formation. Such a low level of star formation is similar to the other HLCs including MBM 53, 54, and 55 complexes.

\subsection{Comparison with HI}

Figure 10 shows the integrated intensity map of H{\small \,I} taken from a Leiden-Dwingeloo H{\small \,I} survey (Hartmann \& Burton 1997) superposed on the integrated intensity of CO. The integrated velocity range is from $-$16 to 0 km s$^{-1}$, corresponding to the velocity range of the $^{12}$CO emission. Because an angular resolution of $\sim$ 30$\arcmin$ is coarser than that of the present CO observations by a factor of $\sim$ 10, we discuss here only the overall comparison between CO and H{\small \,I} distributions. The H{\small \,I} distribution is loop-like and the molecular clouds are distributed nearly along the H{\small \,I} loop.

Figure 11 shows the position-velocity diagram of H{\small \,I} integrated from $-$40$\fdg$5 to $-$39$\fdg$5 and $-$52$\fdg$5 to $-$51$\fdg$5 in Galactic latitude, respectively. The hole like structures can be seen in H{\small \,I}, suggesting that these H{\small \,I} clouds are expanding. These expanding structures in Figures 11(a) and 11(b) correspond to the Galactic northern and southern HI shells shown as the thick dashed (semi) ellipses in Figure 10, respectively, while the expanding motion is not seen in CO (See Figure 3). These two expanding shells are also identified by an infrared radiation (Kiss et al. 2004). From Figure 10, an H{\small \,I} cloud around ($l$, $b$) $\sim$ (109$\degr$, $-$52$\degr$) seems to be located on the left side of the southern expanding shell and the molecular clouds are distributed nearby. It is difficult to distinguish with which H{\small \,I} shell the molecular clouds located around ($l$, $b$) $\sim$ (109$\degr$, $-$52$\degr$) are associated because the H{\small \,I} velocities of the two shells are similar with each other.  The shape of $^{12}$CO and SFD100 $\mu$m radiation around ($l$, $b$) $\sim$ (109$\degr$, $-$52$\degr$) in Figure 2 is also similar to the left side of the expanding shell. If this is true, there may be two expanding structures in the present region.  HD886 (109$\fdg$43, $-$46$\fdg$68) is located near the center of the northern expanding shell, indicating that HD886 may be affecting the northern expanding shell.  The parallax of HD886 has been measured to be 9.79$\pm$0.81 mas (Perryman et al. 1997), corresponding to a distance from the Sun of 102$^{+9}_{-8}$ pc.  The proper motion has also been measured to be $\mu\alpha*$=0$\farcs$0047 yr$^{-1}$, $\mu\delta$=$-$0$\farcs$0082 yr$^{-1}$ by Perryman et al. (1997).  From this proper motion, the velocity of HD886 in the L-B map is estimated to be  $\sim$ 4.3$\times$10$^{-6}$ pc yr$^{-1}$ at $\sim$ 100 pc (see Figure 10). We use typical values of the stellar wind for B2(IV) star on $dM$/$dt$=10$^{-9}$ $M_{\sun}$ yr$^{-1}$ and $V_{\infty}$ = 1000 km s$^{-1}$ (e.g., Snow 1982).  From these parameters, the energy injected to the northern shell is estimated to be $\sim$ 10$^{47}$ ergs in a few $\times$10$^{6}$ yr.  The expanding energy of the northern shell is estimated to be $\sim$ 10$^{48}$ ergs from the atomic and molecular hydrogen, using that the masses of amomic and molecular hydrogen associated with the northern shell are $\sim$ 400 and 42 $M_{\sun}$, respectively, the expanding velocity is $\sim$ 7 km s$^{-1}$ which is estimated from Figure 11 and the equation of $E_{\rm exp}$=1/2$M$$V_{\rm exp}^{2}$.  Since the expanding energy of the atomic and molecular hydrogen is comparable to the energy from HD886, additional source of energy other than HD886 such as photo evaporation is needed to explain the expanding energy because the energy conversion efficiency of the stellar wind is $\lesssim$ 10\%.  The energy of the southern shell is estimated to be $\sim$ 10$^{47}$ ergs, by using that the masses of atomic and molecular hydrogen are $\sim$ 1000 and 16 $M_{\sun}$, respectively, and that the expanding velocity is $\sim$ 9 km s$^{-1}$.  We could not find possible candidates of the energy source for the expanding feature in the literature (SIMBAD) and there are no counterparts in optical or X-ray wavelength. Although we could not identify the possible candidates, we cannot exclude a possibility that these objects may have escaped from the region in a few $\times$ 10$^{6-7}$ yr after forming these structures.

\section{DISCUSSION}

\subsection{Physical States of the Small Clouds}

The present observations have revealed numerous molecular clouds having very small mass of less than 1 $M_{\sun}$.  It is of considerable interest to pursue the physical states of these SCs from view points of cloud physics and chemistry as well as of the origin of molecular clouds.

We shall hereafter focus on the low mass $^{12}$CO clouds whose mass is less than 1 $M_{\sun}$.  The total number of such clouds is 65 among 78. The $^{13}$CO emission has been searched for toward 24 of the 43 $^{12}$CO low-mass clouds whose mass is in a range of 0.1--1.0 $M_{\sun}$ and has been detected from 8 of them.
Figure 12 shows correlations of the molecular column density, estimated from $^{12}$CO and $^{13}$CO, of the clouds where both $^{12}$CO and $^{13}$CO emission are detected.  It is seen that almost all of the $^{12}$CO clouds having molecular column density greater than 5$\times$10$^{20}$ cm$^{-2}$, corresponding to the visual extinction of 0.55 mag if we use the relationship of $N(\rm H_2)$=9.4$\times$10$^{20}$$\times$$A$v cm$^{-2}$  (Bohlin et al. 1978; Hayakawa et al. 1999), show significant $^{13}$CO emission.  We note that there are 22 $^{12}$CO clouds whose mass is less than 0.1 $M_{\sun}$; for those it is doubtful that the $^{13}$CO emission is so significant as those whose mass is a range of 0.1--1.0 $M_{\sun}$ although only 3 of them were searched for the $^{13}$CO emission in the present study. We note that we ignore the possible contribution of atomic hydrogen in the above relationship. If we take into account the distribution of atomic hydrogen as $N$(H{\small \,I}), the visual extinction would increase by 0.2 $\sim$ 0.3 mag. This may be explained as that the $^{13}$CO emitting regions become significant when $A$v becomes larger than $\sim$ 1 mag, marginally enough to shield the ultraviolet radiation to protect $^{13}$CO molecules (e.g., Warin et al. 1996), although $^{13}$CO molecules may be also affected from the ultraviolet radiation because the $N(\rm H_2)$ derived from $^{13}$CO is lower than that derived from $^{12}$CO in most of molecular clouds.

The peak intensity ratio of $^{12}$CO and $^{13}$CO is around 5, much smaller than the terrestrial abundance ratio of 89, indicating that the $^{12}$CO emission is optically thick in the clouds where the both emissions are detected. The maximum $^{12}$CO peak temperature of the present brightest $^{12}$CO emission is 6 K, and this suggests that the excitation temperature is consistent with the kinetic temperature of $\sim$ 10 K, typical to the local dark clouds. The low mass clouds show lower $^{12}$CO peak temperatures down to 1 K, significantly less than the brightest peak intensities. It is not clear if this is due to the lower excitation temperatures or due to smaller filling factors significantly less than 1.  In order to clarify this point we need observations of the present low mass clouds at much higher angular resolutions.

\subsection{Origin of the Very Small Clouds}

  Figure 13 shows the histograms of mass and sizes of $^{12}$CO clouds. In the present region, we have detected a large number of molecular clouds of mass less than 0.1 $M_{\sun}$ and sizes less than 0.1 pc not detected so far in the other regions.  We may ask why molecular clouds of mass less than 0.1 $M_{\sun}$ and size less than 0.1 pc such as the present clouds have not been detected so far. The main reason for this is perhaps the paucity of high-resolution observational studies of nearby molecular clouds. Most of the observations of the local high latitude clouds were made at lower resolutions of 10 arc-min or at a coarse grid spacing of 1 degree, both of which are unable to detect and resolve the present low mass clouds. This suggests that the low mass clouds similar to the present ones may not be uncommon in the interstellar space and warrant more extensive searches for them in the other parts of the sky.

  It is interesting to compare the physical parameters of the present VSCs with theoretical studies. The typical size of these HLCs, $\sim$ 0.1 pc, is significantly less than the Jeans length of 1.3 pc -- 7 pc for molecular gas with $T$=10 K and $n$(H$_{2}$)=10--100 cm$^{-3}$. If we take temperature higher than 10 K, the length becomes even larger. Figure 14 shows the radial distribution of mass surface density which is derived by dividing the mass in circular annulus of a radius by the area of the circular annulus.  Mass surface density in the present region is fairly flat in radius because there is no massive molecular cloud in the center of the VSCs.  On the other hand, mass surface density in the other regions has a gradient for radius, indicating that there are small molecular clouds around a massive molecular cloud whose mass is several dozen $M_{\sun}$ or greater (e.g., Sakamoto 2002, Sakamoto \& Sunada 2003).

In these regions, the gravity of the massive molecular cloud may contribure to the formation of these small clouds by increasing the pressure in the surroundings. 
These suggest that mechanisms other than gravitational instability might contribute to the formation of present VSCs.  A theory of molcular cloud formation is discussed by Koyama \& Inutsuka (2002).  According to them, molecular clouds smaller than the Jeans-length can be formed in the shocked layer through the thermal instability. We shall present some considerations by comparing the observational results with their theoretical results.

Present VSCs are likely to be affected by HD886 in the last few $\times$10$^{6}$ yr (for details, see section 5.2).  Although the mechanical luminosity from HD886 injected to the loop-like structure during a few $\times$10$^{6}$ yr is low to explain the expanding of the interstellar matter, it is a possibility that the stellar wind of HD886 is the source of shock.  Koyama \& Inutsuka (2002) assumed shock velocity of 26 km s$^{-1}$ and density of 0.6 cm$^{-3}$ as an initial pre-shock condition and find that the region of density greater than 100 cm$^{-3}$ grows in size to $\sim$ 0.2 pc $\times$ 0.1 pc in 1.06$\times$10$^{6}$ yr and the internal structure consists of some filaments. The velocity dispersion of CO derived by Koyama \& Inutsuka (2002) is a few km s$^{-1}$. 
The size of the smallest molecular clouds and the velocity dispersion of CO are comparable to those derived from Koyama \& Inutsuka (2002). We cannot resolve the internal structure of the VSCs because the present VSCs are detected with only a few points for each. The typical column density derived by Koyama \& Inutsuka (2002) is 2$\times$10$^{20}$ cm$^{-2}$, while the column density of the VSCs is estimated to be $\sim$ 1.6$\times$10$^{20}$ cm$^{-2}$ (see Figure 4). 
The surface filling factor of the region of density greater than 100 cm$^{-3}$ in Koyama \& Inutsuka (2002) is roughly estimated to be 30--40\%. In this surface filling factor the column density estimated from the observations is consistent with that derived from Koyama \& Inutsuka (2002) and the low temperature of the VSCs is consistent with the result that their peak temperature is lower than that of the typical local dark clouds. These results indicate a potential that the present VSCs are formed in the shock compressed layer through thermal instability. 
In order to compare the observational results with the theoretical simulation on internal temperature and density structure in more details, the observations of higher resolutions are needed.

\section{CONCLUSIONS}

We have made a large-scale survey of high Galactic latitude molecular clouds in the $J$ = 1--0 lines of $^{12}$CO and $^{13}$CO toward a large scale structure located around ($l$, $b$) $\sim$ (109$\degr$, $-$45$\degr$) with NANTEN.  This survey spatially resolved the distribution of molecular gas associated with the large scale structure.  The main conclusions of the present study are summarized as follows:

\begin{enumerate}
\item  The $^{12}$CO observation covered the entire large loop-like structure.  The loop-like structure consits of very small clumpy clouds. The $^{12}$CO clouds are concentrated on the north to north-west of the loop-like structure and toward the south of that.  We identified 78 $^{12}$CO clouds in the observed region.  The total mass is estimated to be $\sim$ 64 $M_{\sun}$ if we assume the conversion factor from CO intensity to $N$(H$_2$) as 1.0$\times$10$^{20}$ cm$^{-2}$/(K km s$^{-1}$).

\item  We performed $^{13}$CO observations in and around the whole area where the peak temperature of $^{12}$CO is more than 2.0 K.  We identified 33 $^{13}$CO clouds and derived physical properties under the assumption of LTE.  

\item  The mass spectra are well fitted by a power law, $dN/dM$ $\propto$ $M^{-1.53\pm0.13}$ for the $^{12}$CO clouds and $dN/dM$ $\propto$ $M^{-1.36\pm0.10}$ for the $^{13}$CO clouds. These spectral indices are similar to those derived in the other regions.

\item  The size and the line width relation of $^{13}$CO clouds is fitted by a least-squares method, log($\Delta V$) = (0.22$\pm$0.43) $\times$ log($R$) + (0.37$\pm$0.52) (c.c.=0.23), but the correlation is not good.

\item  Present $^{13}$CO clouds are far from the virial equilibrium, indicating that $^{13}$CO clouds are not gravitationally bound. $M_{\rm vir}$ and $M_{\rm LTE}$ relation can be fitted by a least-squares method as log($M_{\rm vir}$) = (0.91$\pm$0.30) $\times$ log($M_{\rm LTE}$) + (2.23$\pm$0.29) (c.c.=0.66).  This index is slightly different from the indices in the other regions although the tendency that molecular clouds are more vilialized as the mass increases is consistent with the other regions.

\item  There is no sign of star formation from the comparison of IRAS point sources and Point Source Catalog of Two-Micron All-Sky survey in the present region. This suggests that molecular clouds in this region are not the site of present star formation or the remnants of past star formation.

\item  There may be two expanding shells in the present region as inferred from H{\small \,I}  although we cannot identify them from CO.  The total mechanical luminosity of HD886 during the last few $\times$ 10$^{6}$ yr is comparable to the expanding energy of the northern expanding H{\small \,I} shell.  This indicates that some additional source of energy other than HD886 is needed to explain the expanding energy. 

\item  $^{13}$CO emission is significantly detected in the $^{12}$CO clouds having molecular column density greater than 5$\times$10$^{20}$ cm$^{-2}$. This may be explained as that the $^{13}$CO emitting regions become significant when $A$v becomes larger than $\sim$ 1 mag, marginally enough to shield the ultraviolet radiation to protect $^{13}$CO molecules.

\item  There is a possibility that very small clouds have been formed in the shoked layer through the thermal instability. The stellar wind of HD886 may be the source to creat shocks, forming the loop-like structure where the very small clouds are embedded.

\end{enumerate}

\acknowledgments
We greatly appreciate the hospitality of all staff members of the Las
Campanas Observatory of the Carnegie Institution of Washington. The
NANTEN telescope is operated based on a mutual agreement between
Nagoya University and the Carnegie Institution of Washington. We also
acknowledge that the operation of NANTEN can be realized by
contributions from many Japanese public donators and companies. Three
of the authors (N.M., T.O., and Y.F.) acknowledge financial support
from the scientist exchange program under bilateral agreement between
JSPS (Japan Society for the Promotion of Science) and CONICYT (the
Chilean National Commission for Scientific and Technological
Research). This research has made use of the SIMBAD astronomical database
operated by CDS, Strasbourg, France.  This publication makes use of data products from the Two Micron All Sky Survey, which is a joint project of the University of Massachusetts and the Infrared Processing and Analysis Center/California Institute of Technology, funded by the National Aeronautics and Space Administration and the National Science Foundation.  This research has made use of the IRAS point sources from the NASA/IPAC Infrared Science Archive, which is operated by the Jet Propulsion Laboratory, California Institute of Technology, under contract with the National Aeronautics and Space Administration.

\clearpage

\begin{figure}
\epsscale{0.50}
\plotone{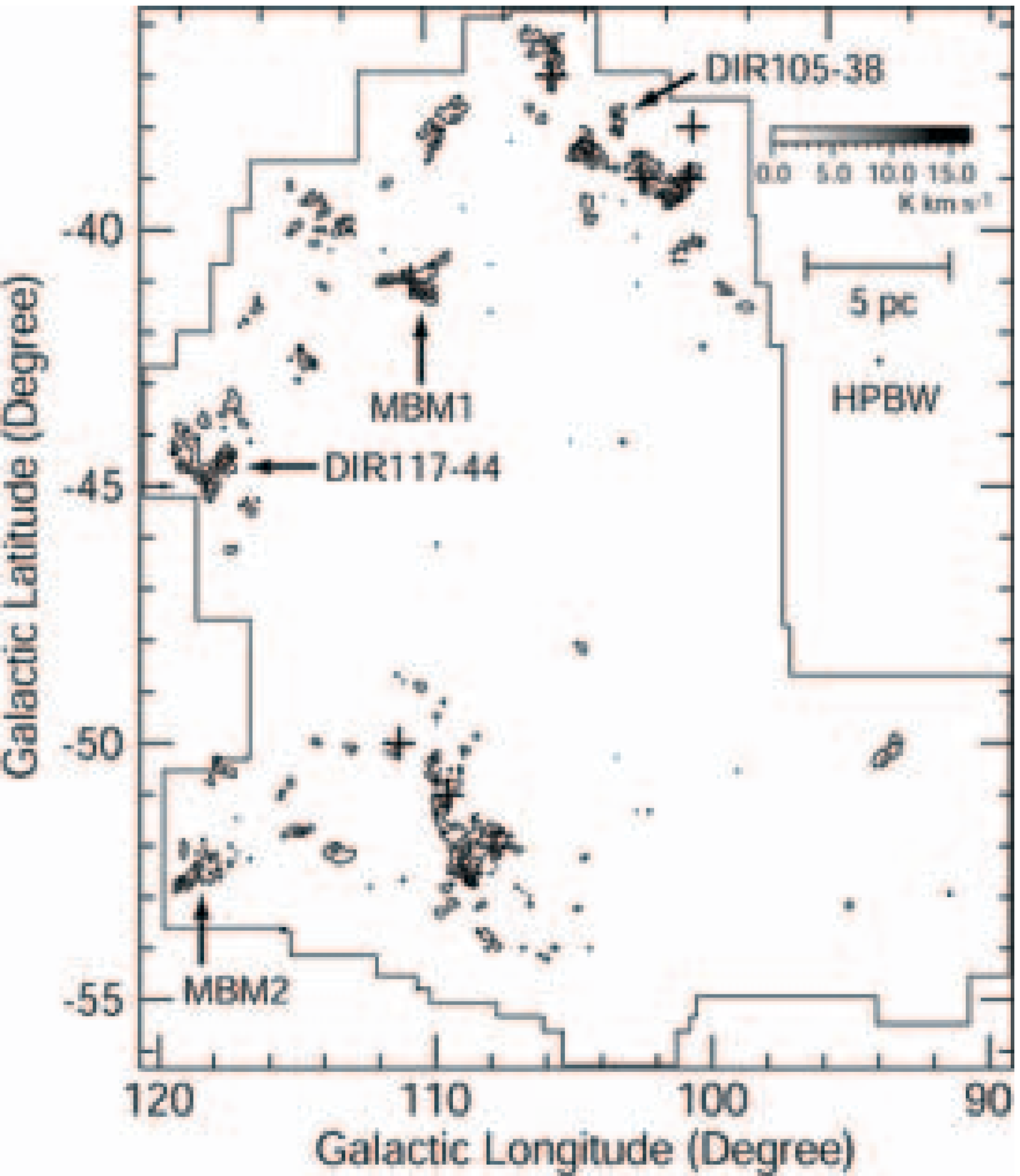}
\caption{Total integrated intensity map of $^{12}$CO ($J$ = 1--0) shown in Galacticcoordinates. The  lowest contour and the separation between contours are
0.77 and 3.08 K km s$^{-1}$, respectively. The solid line represents the
observed area. The crosses and filled circle indicate the position where
Magnani et al. (2000) and (1986) detected $^{12}$CO, respectively.
The locations of DIR clouds in Reach et al. (1998) and MBM clouds in Magnani et
al. (1985) are shown in the figure. \label{fig1}}
\end{figure}

\clearpage

\begin{figure}
\plotone{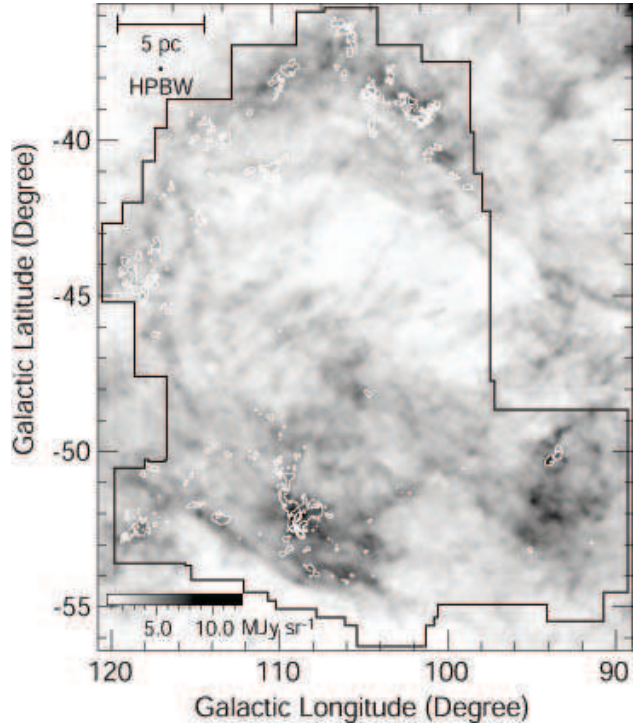}
\caption{Total integrated intensity map of $^{12}$CO ($J$ = 1--0)
superposed on the SFD 100 $\mu$m image derived by Schlegel et al. (1998).
The gray scale shows the intensity of 100 $\mu$m and
white lines represent the integrated intensity of $^{12}$CO emission.
The lowest contour and the separation between contours are the same as
Figure 1.\label{fig2}}
\end{figure}

\clearpage

\begin{figure}
\epsscale{0.50}
\plotone{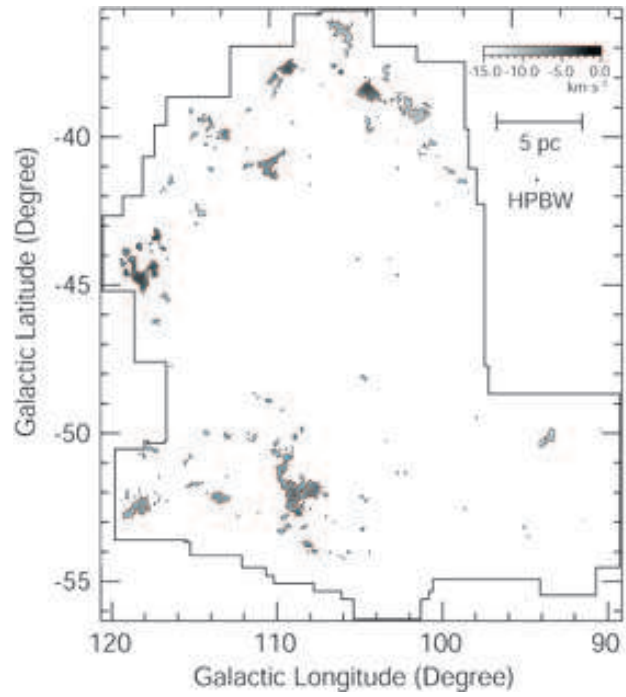}
\caption{The peak radial velocity map of $^{12}$CO ($J$ = 1--0) emission.
The velocity of $^{12}$CO emission is derived from the
single gaussian fitting. Solid lines in the observed area
show the boundary of the $^{12}$CO clouds. \label{fig3}}
\end{figure}

\clearpage

\begin{figure}
\epsscale{0.8}
\plotone{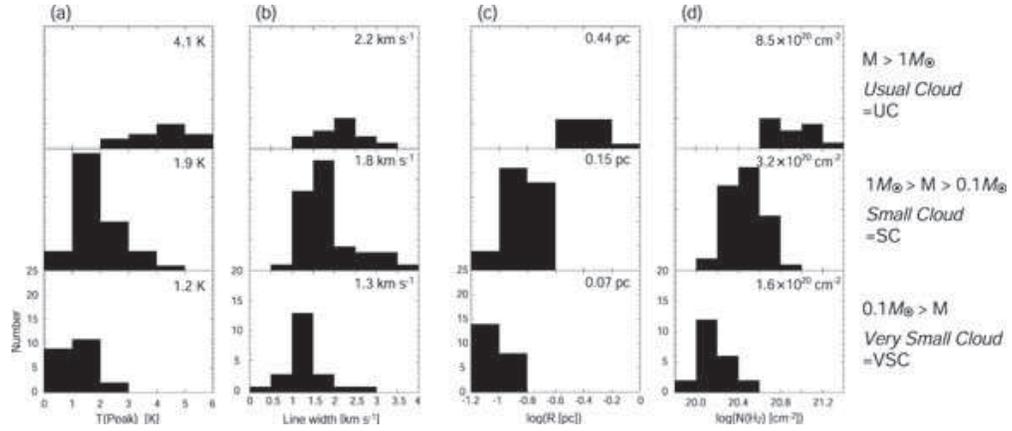}
\caption{Histograms of
(a) Peak temperature, (b) Line width, (c) Radius and (d) Column density of molecular hydrogen on $^{12}$CO clouds.  Histograms of each physical property consist of 3 different ranges on the mass of $^{12}$CO clouds. The range of mass of $^{12}$CO clouds, M, is M $>$ 1 $M_{\sun}$,
1 $M_{\sun}$ $>$ M $>$ 0.1 $M_{\sun}$ and 0.1 $M_{\sun}$ $>$ M from the top panel. The number in each panel is the average value.
\label{fig4}}
\end{figure}

\clearpage

\begin{figure}
\plottwo{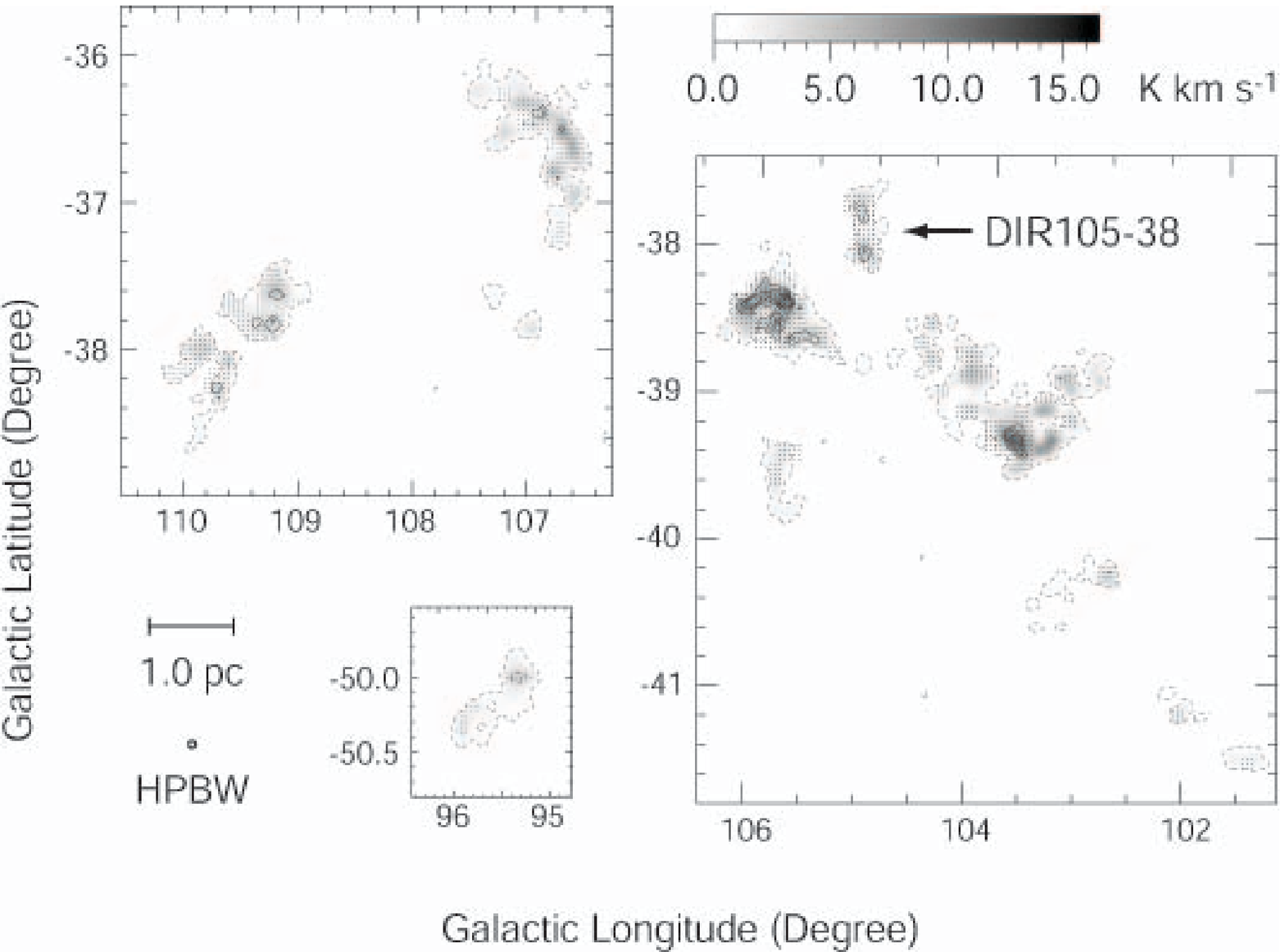}{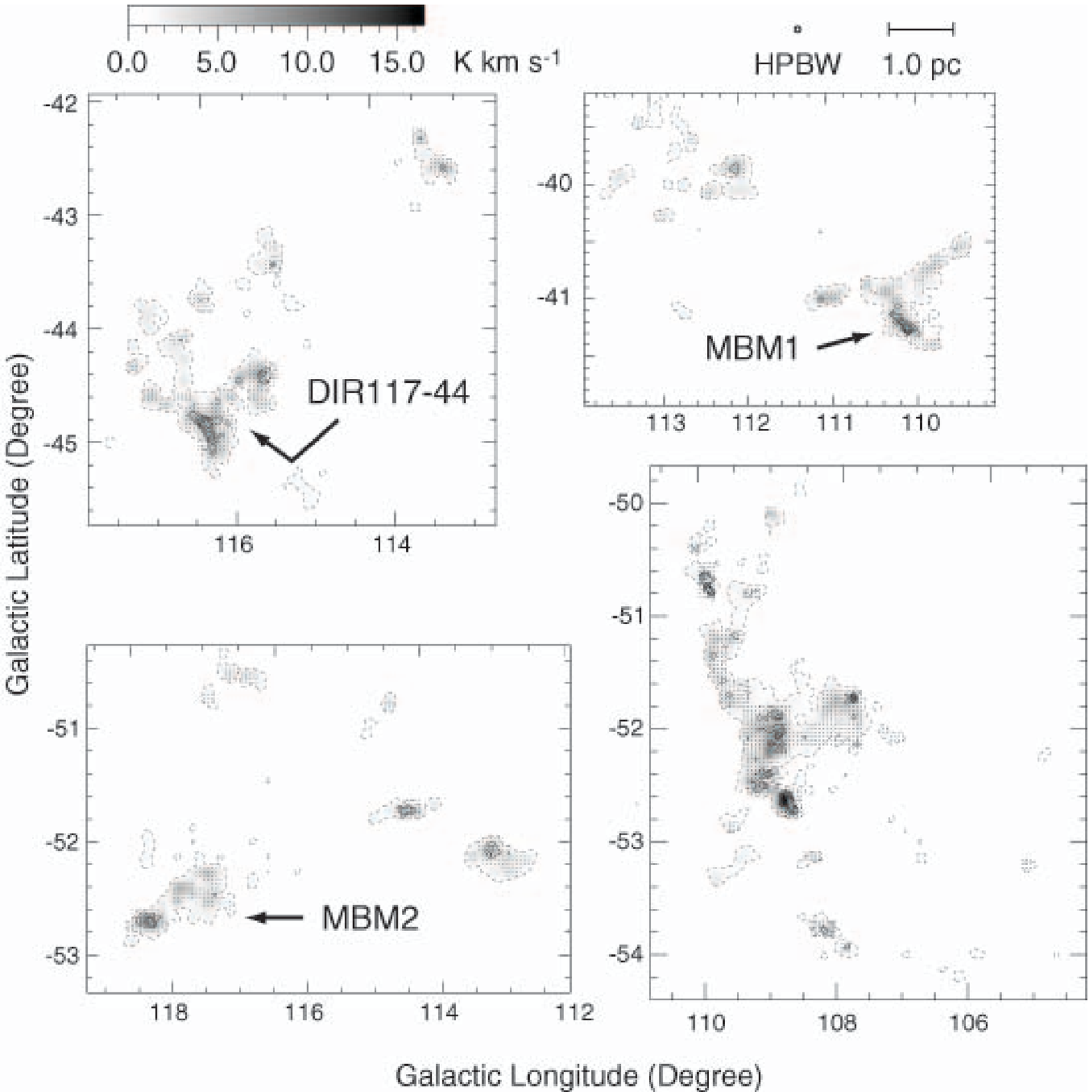}
\caption{Total integrated intensity map of $^{13}$CO ($J$ = 1--0) shown in
Galactic coordinates (bold lines). The lowest contour and the separation
between contours are 0.3 K km s$^{-1}$.
Gray scale shows the velocity integrated
intensity map of $^{12}$CO ($J$ = 1--0). Dashed lines show the boundary
of $^{12}$CO clouds.
Dots indicate the positions observed in $^{13}$CO. The linear scale of all the figures are the same. \label{fig5}}
\end{figure}

\clearpage

\begin{figure}
\plotone{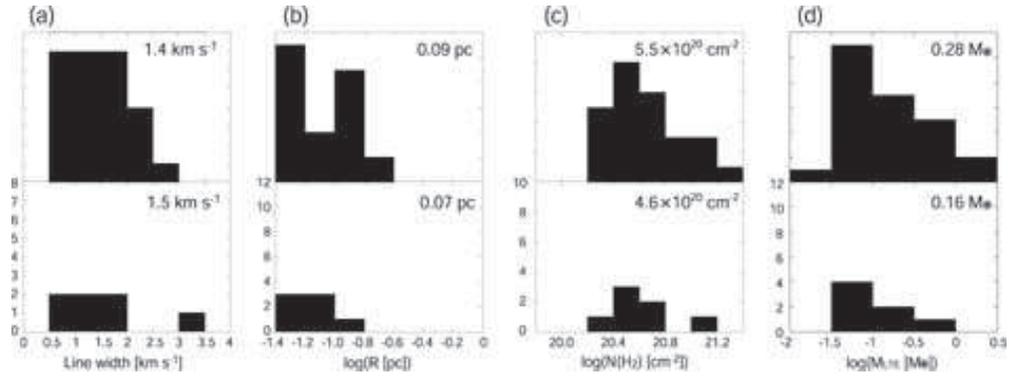}
\caption{Histograms of (a) Line widht, (b) Radius, (c) Column density of molecular hydrogen and (d) LTE mass on the $^{13}$CO clouds. The $^{13}$CO clouds identified in the UC and SC are shown in the upper and lower histogram, respectively.  The range and the interval of the histograms of (a), (b), and (c) are the same
as Figure 4. The number in each panel is the average value. \label{fig6}}
\end{figure}

\clearpage

\begin{figure}
\epsscale{0.70}
\plotone{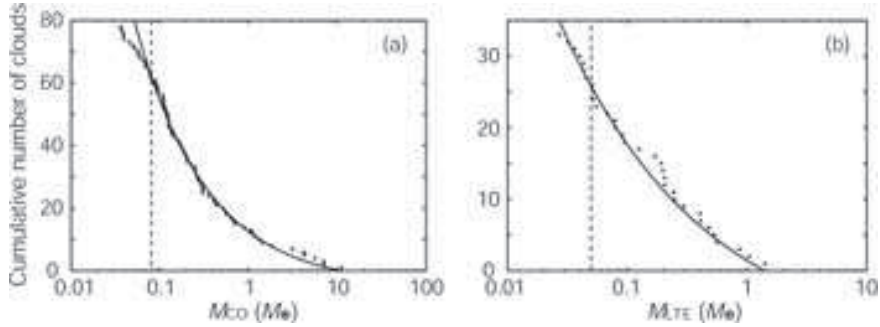}
\caption{(a) Mass spectrum of the $^{12}$CO clouds. The number of clouds,
$N$($\geq$$M_{CO}$), with mass greater than $M_{\rm CO}$ is plotted
against $M_{\rm CO}$,
along with the best-fitting power law (solid lines).  The best fitting is expected as $N$($\geq$$M_{\rm CO}$) =17.33$\times$$M_{\rm CO}$$^{-0.53}$$-$4.78 for the mass range of $\geq$0.08 $M_{\sun}$, which is the completeness limit indicated by the broken line,
derived by using the maximum likelihood method
(Crawford et al. 1970). (b) Same as (a), but for $^{13}$CO clouds with $M_{\rm LTE}$.
The best fitting is expected as
$N$($\geq$$M_{\rm CO}$) =12.54$\times$$M_{\rm LTE}$$^{-0.36}$$-$11.06 for the mass
range of $\geq$0.05 $M_{\sun}$. \label{fig7}}
\end{figure}

\clearpage

\begin{figure}
\epsscale{0.50}
\plotone{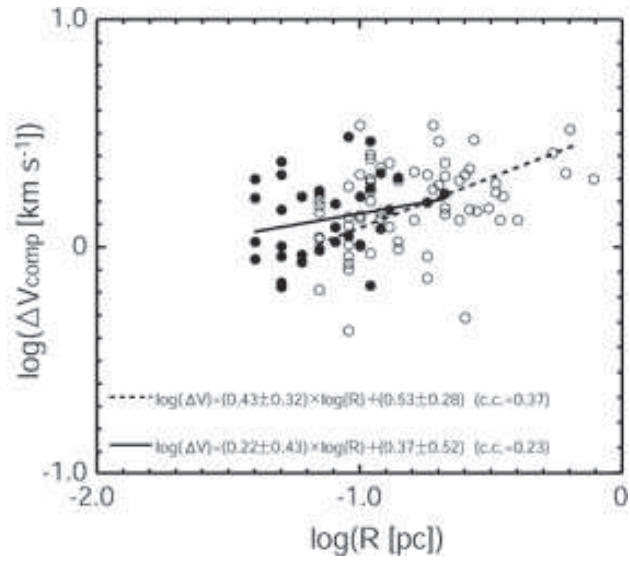}
\caption{Plots of composite line width, $\Delta V$$_{\rm comp}$ versus radius of
$^{13}$CO clouds, R.
The filled and open circles indicate the $^{13}$CO clouds in present region and
those in MBM 53, 54 and 55 complex, respectively. The solid and dashed line indicate the least-squares fit to these plots in present region and MBM 53, 54 and 55, respectively. \label{fig8}}
\end{figure}

\clearpage

\begin{figure}
\epsscale{0.50}
\plotone{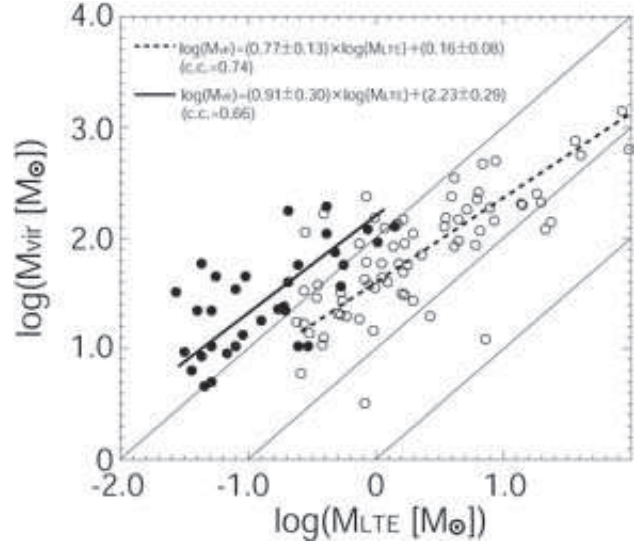}
\caption{Plots of $M_{\rm vir}$ versus $M_{\rm LTE}$. The filled and
open circles indicate the $^{13}$CO cluods in present region and in MBM 53,
54, and 55 complex, respectively. The three thin solid lines represent
$M_{\rm vir}$=10$^{2}$$\times$$M_{\rm LTE}$, $M_{\rm vir}$=10$\times$$M_{\rm LTE}$ and $M_{\rm vir}$=$M_{\rm LTE}$ from the top.
The thick solid and dashed lines represent the least-squares fits to these plots in present region and those MBM 53, 54, and 55 complex, respectively. \label{fig9}}
\end{figure}

\clearpage

\begin{figure}
\plotone{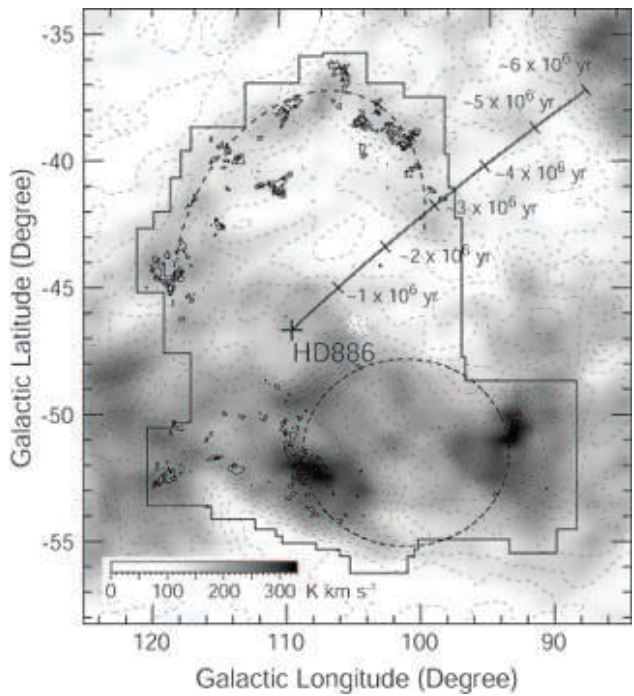}
\caption{Total integrated intensity map of $^{12}$CO ($J$ = 1--0) superposed on
that of H{\small \,I}. The solid contours represent the integrated intensity of $^{12}$CO.  The dashed thin contours and gray scale represent the integrated intensity of H{\small \,I}.
The velocity coverage of H{\small \,I} is $-$16 to 0 km s$^{-1}$, corresponding to that
of $^{12}$CO. The lowest contour and the separation between contours of H{\small \,I} are
25 K km s$^{-1}$ and those of CO are the same as Figure 1.  The thick dashed contours represent
the location of the expanding shells. The cross indicates the position of HD886.  The path of the
proper motion of HD886 is illustrated for the last $\sim$ 6$\times$10$^{6}$ yr by the solid line.
\label{fig10}}
\end{figure}

\clearpage

\begin{figure}
\epsscale{0.70}
\plotone{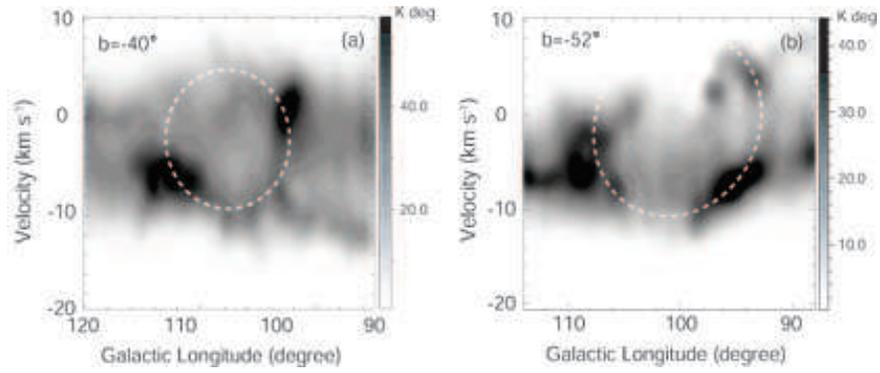}
\caption{Position-velocity map of H{\small \,I} integrated from (a) $-$40$\fdg$5 to $-$39$\fdg$5 in Galactic latitude, (b) $-$52$\fdg$5 to $-$51$\fdg$5 in Galactic latitude. The dashed line in each panel represent the location and the extent of the H{\small \,I} holes drawn by hand.
\label{fig11}}
\end{figure}

\clearpage

\begin{figure}
\epsscale{0.60}
\plotone{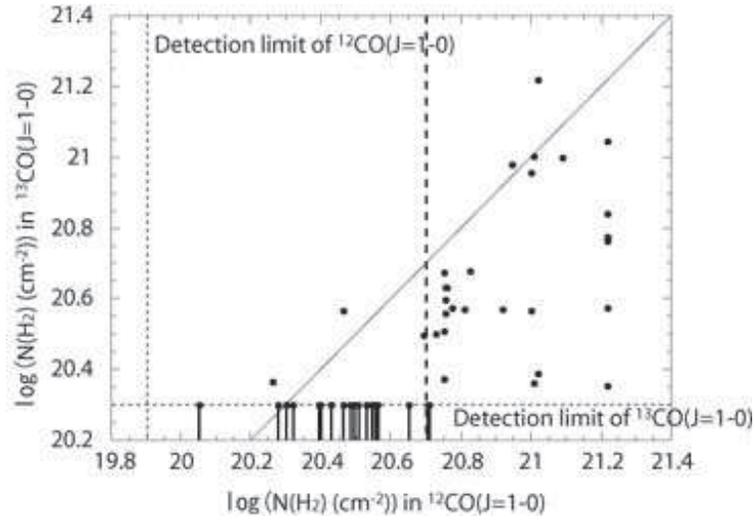}
\caption{Plots of $N$(H$_{2}$) derived from $^{12}$CO versus it from $^{13}$CO at the peak position of
each cloud which were observed in $^{13}$CO emission. The vertical and horizontal dashed lines indicate
the detection limit of the observation of $^{12}$CO and $^{13}$CO, respectively. The thick dotted line indicates
$N(\rm H_2)$ from $^{12}$CO = 5$\times$10$^{20}$ cm$^{-2}$ and the solid line indicates the equal one between
$N(\rm H_2)$ derived from $^{12}$CO and $^{13}$CO.
\label{fig12}}
\end{figure}

\clearpage

\begin{figure}
\epsscale{0.70}
\plotone{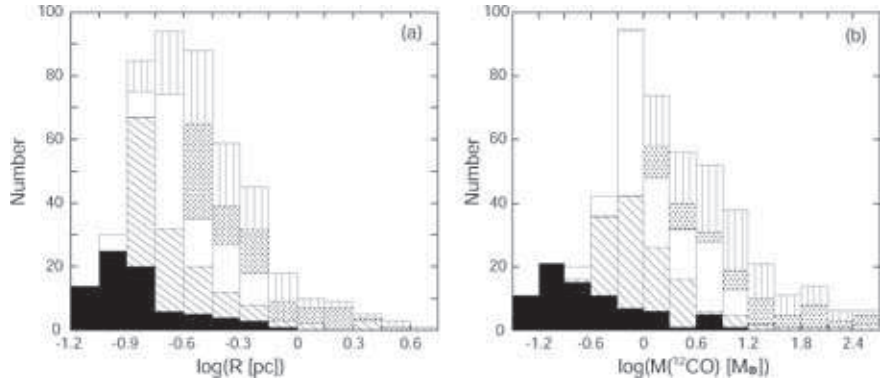}
\caption{Histograms of (a) radii of $^{12}$CO clouds and (b) mass of them.
The dark shaded areas indicate the $^{12}$CO clouds in present region.
The vertical lined, the dotted lined, the light shaded and  the
diagonal lined areas indicate $^{12}$CO clouds
derived by Tachihara et al. (2001), Magnani et al. (1996), Onishi et
al. (2001) and Yamamoto et al. (2003), respectively. \label{fig13}}
\end{figure}

\clearpage

\begin{figure}
\epsscale{0.50}
\plotone{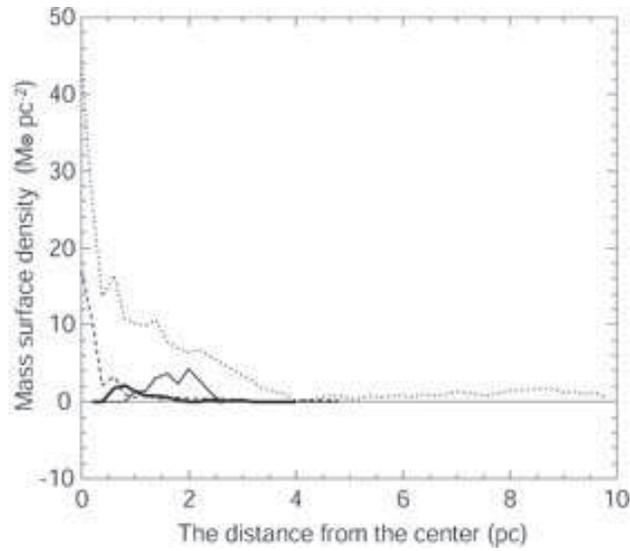}
\caption{The radial distribution of mass surface density. The thick and thin lines indicate the
regions around (113$\degr$, $-$51$\degr$) and (113$\degr$, $-$40$\degr$) in Galactic coordinates in
present region, respectively. The dashed and dotted lines indicate the regions around HLCG92$-$35
and MBM 53 (Yamamoto et al. 2003), respectively.
\label{fig14}}
\end{figure}

\clearpage

\catcode`@=\active \def@{\phantom{0}}
\begin{deluxetable}{ccccccccc}
\tablewidth{0pt}
\tablecaption{Physical Properties of $^{12}$CO Clouds \label{tbl-1}}
\tablehead{
\colhead{No.} & 
\colhead{$l$} & 
\colhead{$b$} & 
\colhead{$T_{\rm R}^{\ast}$} & 
\colhead{$\Delta V$} &
\colhead{$V_{\rm LSR}$} & 
\colhead{$R$} & 
\colhead{$N$(H$_2$)} &
\colhead{$M_{\rm CO}$} \\
\colhead{(1)} &
\colhead{(2)} &
\colhead{(3)} &
\colhead{(4)} &
\colhead{(5)} &
\colhead{(6)} &
\colhead{(7)} &
\colhead{(8)} &
\colhead{(9)}
}
\startdata
@1 & @95.52 & $-$50.00 & 3.2 & 1.3 & $-$10.7 & 0.31 & @4.7 & @1.07 \\
@2 & @96.11 & $-$53.13 & 1.1 & 1.3 & @$-$9.4 & 0.09 & @0.8 & @0.07 \\
@3 & 101.43 & $-$41.53 & 1.5 & 1.5 & @$-$9.5 & 0.17 & @1.9 & @0.27 \\
@4 & 101.91 & $-$41.20 & 0.7 & 0.5 & @$-$9.5 & 0.08 & @0.9 & @0.04 \\
@5 & 102.09 & $-$41.20 & 1.9 & 1.4 & @$-$8.9 & 0.10 & @3.4 & @0.12 \\
@6 & 102.19 & $-$41.07 & 0.9 & 1.1 & @$-$8.9 & 0.08 & @1.0 & @0.04 \\
@7 & 102.80 & $-$40.27 & 2.5 & 1.6 & @$-$8.4 & 0.16 & @4.5 & @0.30 \\
@8 & 103.32 & $-$40.33 & 1.1 & 1.3 & @$-$8.3 & 0.11 & @1.6 & @0.11 \\
@9 & 103.47 & $-$40.60 & 1.0 & 1.2 & @$-$7.3 & 0.08 & @1.0 & @0.04 \\
10 & 103.74 & $-$39.33 & 5.2 & 2.9 & $-$10.1 & 0.59 & 12.3 & @7.02 \\
11 & 104.20 & $-$38.93 & 0.8 & 1.0 & $-$15.7 & 0.10 & @2.0 & @0.09 \\
12 & 104.48 & $-$38.53 & 1.4 & 3.5 & @$-$8.6 & 0.16 & @4.5 & @0.44 \\
13 & 104.81 & $-$38.80 & 1.4 & 1.2 & @$-$5.0 & 0.10 & @1.5 & @0.08 \\
14 & 105.07 & $-$38.80 & 1.1 & 1.4 & @$-$5.7 & 0.08 & @1.8 & @0.06 \\
15 & 105.08 & $-$52.27 & 1.0 & 1.2 & @$-$0.1 & 0.07 & @1.2 & @0.04 \\
16 & 105.10 & $-$38.07 & 2.5 & 2.4 & $-$12.7 & 0.26 & @5.7 & @1.19 \\
17 & 105.22 & $-$53.20 & 1.2 & 1.3 & @$-$3.7 & 0.09 & @1.1 & @0.05 \\
18 & 105.40 & $-$48.13 & 1.6 & 1.8 & @$-$9.0 & 0.12 & @3.0 & @0.20 \\
19 & 105.77 & $-$38.40 & 5.6 & 1.9 & @$-$5.0 & 0.47 & 10.5 & @5.60 \\
20 & 105.80 & $-$39.60 & 1.7 & 1.8 & @$-$9.8 & 0.24 & @3.1 & @0.70 \\
21 & 105.82 & $-$54.00 & 1.1 & 1.4 & @$-$4.8 & 0.07 & @1.3 & @0.04 \\
22 & 106.76 & $-$36.53 & 2.7 & 3.0 & @$-$9.8 & 0.53 & @8.3 & @4.29 \\
23 & 106.97 & $-$37.87 & 1.7 & 1.7 & @$-$1.2 & 0.12 & @3.4 & @0.17 \\
24 & 107.32 & $-$37.60 & 1.2 & 1.5 & @$-$2.3 & 0.12 & @1.6 & @0.11 \\
25 & 107.38 & $-$52.00 & 1.0 & 2.6 & @$-$7.5 & 0.12 & @2.1 & @0.13 \\
26 & 107.87 & $-$53.93 & 0.8 & 1.6 & @$-$4.9 & 0.11 & @1.8 & @0.11 \\
27 & 108.21 & $-$53.73 & 2.2 & 1.4 & @$-$5.1 & 0.16 & @2.9 & @0.25 \\
28 & 108.33 & $-$53.13 & 2.2 & 1.2 & @$-$0.2 & 0.07 & @3.5 & @0.08 \\
29 & 108.78 & $-$52.60 & 5.7 & 2.4 & @$-$6.6 & 0.79 & 16.6 & 11.13 \\
30 & 109.00 & $-$52.13 & 3.3 & 1.9 & @$-$1.5 & 0.31 & @5.1 & @1.43 \\
31 & 109.00 & $-$50.07 & 2.5 & 1.6 & @$-$4.7 & 0.13 & @5.1 & @0.26 \\
32 & 109.11 & $-$50.60 & 0.9 & 2.2 & @$-$5.9 & 0.09 & @1.6 & @0.06 \\
33 & 109.17 & $-$37.60 & 4.0 & 1.5 & @$-$4.4 & 0.36 & @5.6 & @1.83 \\
34 & 109.55 & $-$52.87 & 1.3 & 2.0 & @$-$7.1 & 0.14 & @2.5 & @0.16 \\
35 & 109.68 & $-$38.27 & 4.6 & 1.4 & @$-$8.1 & 0.17 & @5.4 & @0.52 \\
36 & 109.76 & $-$38.00 & 2.7 & 2.0 & @$-$4.9 & 0.21 & @4.7 & @0.63 \\
37 & 109.78 & $-$53.33 & 2.8 & 1.5 & @$-$6.4 & 0.18 & @2.6 & @0.30 \\
38 & 109.84 & $-$50.73 & 4.0 & 2.6 & @$-$7.8 & 0.18 & @5.0 & @0.40 \\
39 & 109.85 & $-$38.60 & 0.8 & 1.8 & @$-$8.1 & 0.12 & @2.1 & @0.13 \\
40 & 110.05 & $-$50.40 & 2.1 & 3.1 & @$-$7.4 & 0.10 & @3.6 & @0.14 \\
41 & 110.06 & $-$41.27 & 4.4 & 2.9 & @$-$6.1 & 0.47 & 10.2 & @4.21 \\
42 & 110.32 & $-$48.93 & 2.9 & 0.9 & @$-$4.2 & 0.12 & @2.6 & @0.17 \\
43 & 110.94 & $-$41.00 & 1.3 & 3.4 & @$-$6.3 & 0.17 & @5.7 & @0.67 \\
44 & 110.98 & $-$39.13 & 1.2 & 2.2 & @$-$9.6 & 0.12 & @2.9 & @0.20 \\
45 & 111.08 & $-$50.13 & 1.6 & 1.0 & @$-$7.2 & 0.13 & @1.7 & @0.11 \\
46 & 112.04 & $-$39.87 & 3.4 & 2.3 & @$-$5.4 & 0.26 & @6.4 & @1.11 \\
47 & 112.40 & $-$40.07 & 1.3 & 2.6 & @$-$7.2 & 0.10 & @3.6 & @0.19 \\
48 & 112.55 & $-$39.60 & 1.2 & 1.9 & @$-$3.6 & 0.16 & @3.0 & @0.29 \\
49 & 112.64 & $-$50.07 & 0.7 & 1.5 & @$-$6.5 & 0.14 & @1.7 & @0.14 \\
50 & 112.72 & $-$41.13 & 1.2 & 1.9 & @$-$6.2 & 0.11 & @2.4 & @0.13 \\
51 & 112.72 & $-$41.13 & 1.2 & 1.9 & @$-$6.2 & 0.11 & @2.4 & @0.13 \\
52 & 112.72 & $-$39.67 & 1.2 & 2.6 & @$-$5.3 & 0.10 & @1.9 & @0.09 \\
53 & 112.84 & $-$40.27 & 1.3 & 1.7 & @$-$5.1 & 0.08 & @2.9 & @0.11 \\
54 & 113.15 & $-$39.47 & 1.5 & 1.2 & @$-$4.7 & 0.16 & @2.5 & @0.23 \\
55 & 113.23 & $-$52.07 & 4.4 & 1.3 & @$-$7.3 & 0.30 & @5.7 & @1.32 \\
56 & 113.42 & $-$42.33 & 3.4 & 2.4 & $-$10.5 & 0.23 & @5.9 & @0.89 \\
57 & 113.47 & $-$39.20 & 0.5 & 1.9 & @$-$5.7 & 0.10 & @1.4 & @0.08 \\
58 & 113.56 & $-$49.93 & 1.3 & 0.9 & @$-$7.5 & 0.12 & @1.1 & @0.09 \\
59 & 113.94 & $-$51.67 & 1.4 & 1.6 & @$-$9.7 & 0.07 & @3.2 & @0.07 \\
60 & 114.38 & $-$51.73 & 3.9 & 1.6 & @$-$9.7 & 0.17 & @6.7 & @0.55 \\
61 & 114.49 & $-$50.80 & 1.5 & 1.8 & @$-$7.8 & 0.09 & @3.2 & @0.12 \\
62 & 114.52 & $-$41.53 & 2.2 & 1.1 & @$-$7.8 & 0.15 & @2.1 & @0.20 \\
63 & 114.83 & $-$51.07 & 0.8 & 1.8 & @$-$8.1 & 0.10 & @1.4 & @0.07 \\
64 & 114.90 & $-$41.73 & 1.5 & 1.2 & @$-$8.3 & 0.08 & @1.1 & @0.04 \\
65 & 115.10 & $-$43.80 & 1.6 & 1.0 & @$-$2.1 & 0.11 & @1.5 & @0.09 \\
66 & 115.16 & $-$45.33 & 1.2 & 1.6 & @$-$8.4 & 0.17 & @1.6 & @0.25 \\
67 & 115.24 & $-$43.40 & 1.8 & 2.0 & @$-$3.5 & 0.23 & @2.7 & @0.57 \\
68 & 115.63 & $-$43.60 & 0.8 & 1.0 & @$-$3.3 & 0.08 & @1.6 & @0.06 \\
69 & 115.74 & $-$46.20 & 1.5 & 1.0 & @$-$6.8 & 0.13 & @2.3 & @0.18 \\
70 & 116.20 & $-$43.73 & 3.2 & 1.5 & @$-$2.6 & 0.16 & @3.6 & @0.31 \\
71 & 116.33 & $-$44.80 & 4.2 & 2.3 & @$-$3.9 & 0.59 & 10.0 & @7.13 \\
72 & 116.45 & $-$50.53 & 2.3 & 1.1 & @$-$7.3 & 0.17 & @3.5 & @0.35 \\
73 & 116.64 & $-$52.33 & 0.9 & 1.5 & @$-$7.6 & 0.07 & @2.1 & @0.05 \\
74 & 116.86 & $-$43.87 & 0.8 & 3.7 & @$-$4.7 & 0.18 & @2.9 & @0.41 \\
75 & 117.01 & $-$50.73 & 2.7 & 0.8 & @$-$7.5 & 0.10 & @2.0 & @0.10 \\
76 & 117.11 & $-$44.33 & 2.3 & 1.1 & @$-$2.6 & 0.14 & @5.1 & @0.25 \\
77 & 118.12 & $-$52.13 & 1.1 & 1.1 & @$-$6.7 & 0.12 & @1.8 & @0.12 \\
78 & 118.23 & $-$52.67 & 4.2 & 2.0 & @$-$7.8 & 0.43 & @8.8 & @3.12 \\
\enddata

\tablecomments{Col. (1) : Cloud number, Col. (2)--(3) : Cloud
peak ($l,b$) position in degree, Col. (4) : Peak temperature in K, Col. (5)
: Line
width of the composite spectrum in km s$^{-1}$, Col. (6) : Peak velocity of the
composite spectrum in km s$^{-1}$, Col. (7) : Radius of the molecular cloud
in pc, Col. (8) : Column density of peak position in 10$^{20}$ cm$^{-2}$,
Col. (9) : Mass of the molecular cloud in $M_{\sun}$. Col. (4) to (6) are derived
by using a single Gaussiun fitting.
}

\end{deluxetable}

\clearpage

\begin{deluxetable}{ccccccccccc}
\tablewidth{0pc}
\tablecaption{Physical Properties of $^{13}$CO Clouds} 
\tablehead
{
\\ 
\colhead{No.}  & 
\colhead{$l$}  &  
\colhead{$b$}  &  
\colhead{$T_{\rm R}^{\ast}$}  &  
\colhead{$\Delta V$}  &  
\colhead{$V_{\rm LSR}$}  &  
\colhead{$R$} &
\colhead{$\tau(^{13}\rm CO$)}  &  
\colhead{$N(\rm H_2)$} &
\colhead{$M_{\rm LTE}$}  &  
\colhead{$M_{\rm vir}$}  \\
\colhead{(1)} &
\colhead{(2)} &
\colhead{(3)} &
\colhead{(4)} &
\colhead{(5)} &
\colhead{(6)} &
\colhead{(7)} &
\colhead{(8)} &
\colhead{(9)} &
\colhead{(10)} &
\colhead{(11)} 
}
\startdata
16a & 105.11 & $-$38.03 & 0.7 & 1.7 & $-$11.4 & 0.06 & 0.11 & @4.0 & 0.08 & @34.6 \\
16b & 105.12 & $-$37.80 & 0.7 & 1.0 & $-$13.1 & 0.07 & 0.12 & @3.6 & 0.09 & @13.5 \\
19a & 105.54 & $-$38.63 & 0.8 & 1.1 & @$-$5.5 & 0.09 & 0.14 & @2.5 & 0.17 & @23.2 \\
19b & 105.77 & $-$38.37 & 2.3 & 1.7 & @$-$4.0 & 0.21 & 0.46 & 16.6 & 1.41 & 129.8 \\
20@ & 103.70 & $-$39.33 & 1.2 & 1.5 & @$-$9.7 & 0.13 & 0.22 & 10.0 & 0.56 & @57.9 \\
22@ & 106.93 & $-$36.40 & 0.7 & 2.1 & @$-$9.4 & 0.05 & 0.13 & @3.7 & 0.06 & @45.2 \\
26@ & 107.87 & $-$53.93 & 0.3 & 1.8 & @$-$7.7 & 0.07 & 0.13 & @2.3 & 0.09 & @46.4 \\
27@ & 108.10 & $-$53.77 & 0.8 & 3.1 & @$-$6.2 & 0.09 & 0.13 & @3.7 & 0.21 & 176.1 \\
29a & 107.82 & $-$51.70 & 1.2 & 1.6 & @$-$5.0 & 0.08 & 0.21 & @7.0 & 0.21 & @40.7 \\
29b & 108.78 & $-$52.63 & 1.4 & 2.0 & @$-$4.6 & 0.14 & 0.26 & 11.1 & 0.87 & 120.6 \\
29c & 108.84 & $-$52.03 & 1.2 & 0.7 & @$-$6.2 & 0.11 & 0.21 & @5.9 & 0.29 & @10.6 \\
29d & 108.89 & $-$52.17 & 1.3 & 1.1 & @$-$6.9 & 0.08 & 0.23 & @3.8 & 0.13 & @18.4 \\
29e & 109.00 & $-$52.40 & 1.1 & 2.1 & @$-$7.8 & 0.12 & 0.20 & @6.0 & 0.41 & 110.6 \\
29f & 109.00 & $-$52.13 & 0.9 & 1.6 & @$-$1.9 & 0.04 & 0.15 & @5.8 & 0.05 & @22.5 \\
29g & 109.16 & $-$51.87 & 1.0 & 1.7 & @$-$5.8 & 0.10 & 0.17 & @2.3 & 0.25 & @57.6 \\
29h & 109.53 & $-$51.17 & 0.8 & 0.9 & @$-$6.5 & 0.05 & 0.15 & @2.9 & 0.04 & @@8.7 \\
29i & 109.64 & $-$51.57 & 0.3 & 2.4 & @$-$7.0 & 0.05 & 0.12 & @3.2 & 0.04 & @59.7 \\
29j & 109.80 & $-$51.37 & 0.4 & 1.5 & @$-$6.7 & 0.05 & 0.14 & @2.5 & 0.04 & @22.3 \\
33a & 109.17 & $-$37.60 & 0.8 & 1.0 & @$-$4.4 & 0.05 & 0.13 & @2.4 & 0.05 & @10.7 \\
33b & 109.21 & $-$37.80 & 0.9 & 0.9 & @$-$4.1 & 0.06 & 0.16 & @4.7 & 0.08 & @10.6 \\
33c & 109.34 & $-$37.80 & 0.7 & 0.7 & @$-$4.5 & 0.05 & 0.13 & @3.2 & 0.05 & @@4.7 \\
35@ & 109.72 & $-$38.27 & 0.9 & 0.7 & @$-$8.1 & 0.05 & 0.16 & @3.2 & 0.05 & @@5.1 \\
38@ & 109.90 & $-$50.73 & 1.1 & 1.8 & @$-$7.7 & 0.11 & 0.19 & 10.5 & 0.48 & @75.3 \\
41a & 110.11 & $-$41.27 & 1.0 & 2.9 & @$-$5.6 & 0.11 & 0.23 & 10.1 & 0.41 & 197.4 \\
41b & 110.19 & $-$41.07 & 0.6 & 2.0 & @$-$7.6 & 0.04 & 0.10 & @2.3 & 0.03 & @33.1 \\
43@ & 111.12 & $-$41.00 & 0.7 & 1.1 & @$-$5.6 & 0.04 & 0.12 & @4.3 & 0.03 & @@9.4 \\
46@ & 112.08 & $-$39.87 & 0.8 & 0.9 & @$-$5.1 & 0.06 & 0.14 & @3.7 & 0.07 & @@9.3 \\
55@ & 113.17 & $-$52.03 & 1.2 & 0.7 & @$-$7.2 & 0.11 & 0.22 & @4.3 & 0.24 & @10.6 \\
56@ & 113.17 & $-$42.60 & 0.8 & 0.9 & $-$10.9 & 0.04 & 0.14 & @3.7 & 0.04 & @@6.5 \\
60@ & 114.32 & $-$51.70 & 0.8 & 1.2 & @$-$9.7 & 0.08 & 0.15 & @4.8 & 0.19 & @24.9 \\
71a & 115.49 & $-$44.40 & 0.9 & 1.0 & @$-$3.5 & 0.10 & 0.16 & @3.7 & 0.20 & @22.2 \\
71b & 116.21 & $-$44.97 & 1.2 & 1.6 & @$-$3.9 & 0.18 & 0.22 & @9.1 & 1.04 & @93.9 \\
78@ & 118.19 & $-$52.70 & 1.5 & 1.2 & @$-$8.1 & 0.12 & 0.28 & @9.6 & 0.53 & @36.7 \\
\enddata 
\begin{flushleft}
{\footnotesize Note---Col. (1) : Cloud number of $^{13}$CO taken from that of $^{12}$CO cloud with
which the $^{13}$CO cloud is associated. If plural $^{13}$CO clouds are associated with one $^{12}$CO cloud,
a sequential alphabet is added, Col. (2)--(3) : Cloud peak ($l$, $b$) position in degree, Col. (4) : Peak temperature in K, Col. (5) : Line
width of the composite spectrum in km s$^{-1}$, Col. (6) : Peak velocity of the
composite spectrum in km s$^{-1}$, Col. (7) : Radius of the molecular cloud in pc, 
Col. (8) :
Optical depth of $^{13}$CO, Col. (9) : Column density of peak position in 10$^{20}$
 cm$^{-2}$,
Col. (10) : Mass of the molecular cloud assuming the LTE in $M_{\sun}$, Col. (11) :
Virial mass of the molecular cloud in $M_{\sun}$. Col. (4) to (6) are derived by using
 a single Gaussiun fitting.}
\end{flushleft} 

\end{deluxetable}

\end{document}